\documentclass[12pt]{article}
\usepackage{amsfonts,amsthm,amscd,mathrsfs,helvet}
\usepackage{amsbsy}
\usepackage[format=hang]{caption}
\usepackage{epsfig,amssymb,amsmath,graphicx,subcaption,verbatim,hyperref,xcolor,
ulem,epstopdf,psfrag,pstool,braket,array,enumerate,multirow,wrapfig}
\usepackage{graphics,color}
\newcommand{\be}{\begin{equation}}
\newcommand{\ee}{\end{equation}}
\newcommand{\bea}{\begin{eqnarray}}
\newcommand{\eea}{\end{eqnarray}}
\newcommand{\bean}{\begin{eqnarray*}}
\newcommand{\eean}{\end{eqnarray*}}
\font\upright=cmu10 scaled\magstep1
\font\sans=cmss12
\newcommand{\ssf}{\sans}
\newcommand{\stroke}{\vrule height8pt width0.4pt depth-0.1pt}
\newcommand{\Z}{\hbox{\upright\rlap{\ssf Z}\kern 2.7pt {\ssf Z}}}

\newcommand{\C}{{\rlap{\rlap{C}\kern 3.8pt\stroke}\phantom{C}}}
\newcommand{\R}{\hbox{\upright\rlap{I}\kern 1.7pt R}}
\newcommand{\CP}{\C{\upright\rlap{I}\kern 1.5pt P}}
\newcommand{\identity}{{\upright\rlap{1}\kern 2.0pt 1}}
\newcommand{\half}{\frac{1}{2}}
\newcommand{\pr}{\partial}

\newcommand{\z}{{\bar z}}

\newcommand{\w}{{\bar w}}
\newcommand{\bphi}{{\bar \phi}}
\newcommand{\mg}{\mathfrak{g}}

\begin{document}
\pagestyle{plain}

\title{\vskip -70pt
\begin{flushright}
{\normalsize DAMTP-2016-84} \\
\end{flushright}
\vskip 50pt
{\bf \LARGE \bf Five Vortex Equations}
 \vskip 30pt
}
\author{{\bf Nicholas S. Manton}
\thanks{email: N.S.Manton@damtp.cam.ac.uk} \\[15pt]
{\normalsize
{\sl Department of Applied Mathematics and Theoretical Physics,}}\\
{\normalsize {\sl University of Cambridge,}}\\
{\normalsize {\sl Wilberforce Road, Cambridge CB3 0WA, England.}}\\
}
\vskip 20pt
\date{December 2016}
\maketitle
\vskip 20pt

\begin{abstract}
The Taubes equation for Abelian Higgs vortices is
generalised to five distinct U(1) vortex equations. These include the
Popov and Jackiw--Pi vortex equations, and two further equations. The Baptista
metric, a conformal rescaling of the background metric by the
squared Higgs field, gives insight into these vortices, and shows that
vortices can be interpreted as conical singularities superposed on the
background geometry. When the background has a constant curvature adapted to 
the vortex type, then the vortex equation is integrable by a 
reduction to Liouville's equation, and the Baptista metric has a
constant curvature too, apart from its conical singularities. The 
conical geometry is fairly easy to visualise in some cases.

\end{abstract}

\vskip 30pt

\newpage

\section{Introduction}

There are a number of vortex equations \cite{JT,Yan,MS}, defined on surfaces 
of suitable curvature, that can be explicitly solved. We say that these
vortex equations are integrable. The known examples are (i) the Taubes
equation of the Abelian Higgs model, defined on a hyperbolic surface
of constant negative curvature \cite{Wit,MR,MM}, (ii) the Popov 
vortex equation defined on a sphere of constant positive curvature 
\cite{Pop,Man}, and (iii) the Jackiw--Pi vortex equation defined on 
a flat plane or torus \cite{JP1,JP2,HZ}. 
In this paper, we investigate more systematically how these integrable 
vortex equations arise, and discover that there are really five 
examples. The first new example we call (iv) the Bradlow vortex 
equation -- it is a reinterpretation of the Bradlow limit \cite{Bra} of
the Taubes equation for vortices on a compact hyperbolic 
surface, where the vortex number saturates its upper 
bound. The second new example is (v) a vortex equation defined 
on a hyperbolic surface, generalising an equation found by Ambj{\o}rn
and Olesen \cite{AO1,AO2}.

Our approach builds on the geometric insights of Baptista \cite{Bap}, who
interpreted vortices on a smooth surface in terms of a new metric -- the
Baptista metric -- which is a conformal rescaling of the background 
metric. If the background metric is $ds_0^2$ and the vortex Higgs field 
is $\phi$, then Baptista's metric is $ds^2 = |\phi|^2ds_0^2$. This
metric is not smooth. It has singularities at the vortex centres, 
where $\phi = 0$. For a vortex of unit winding, the metric
has a conical singularity with cone angle $4\pi$. The construction of
integrable vortices is then closely related to the purely geometrical
problem of constructing surface metrics with given curvature and
conical singularities. This problem has been studied, in particular,
by Troyanov \cite{Tro}.

We do not solve in full generality the five vortex equations, but 
summarise solutions that are known, and find some new ones.
Solutions are obtained using local holomorphic mappings between 
surfaces. Sometimes these maps are globally defined, and can be given
explicitly. This method may generate all solutions,
but not always, and it appears that further solutions must be 
constructed by patching local holomorphic maps together, with a twist. 

We describe in some detail the intrinsic Baptista geometry of a number
of vortices, in the integrable cases. That is, we describe the
geometry of the Baptista metric $ds^2$, without splitting it into its factors 
$ds_0^2$ and $|\phi|^2$. We also note that this metric is an
Einstein metric with conical singularities, in the presence of a cosmological 
constant \cite{DJtH,DJ}. From this perspective, vortices become 
point particles of negative mass, quite different from the usual 
insight that they are smooth solitons on a smooth background surface.

\section{Abelian Higgs Vortices}

All the vortex equations considered here are variants of the first-order 
Bogomolny equations \cite{Bog} of the Abelian Higgs model, which we
review first. These equations model critically coupled vortices that 
neither attract nor repel each other, so the vortices are static, 
2-dimensional soliton or multi-soliton solutions, occupying a flat or curved 
surface. 

More precisely, the equations are defined on a Riemann surface $M_0$
having a metric compatible with its complex structure. $M_0$ 
may be compact, or open with a boundary at infinity. Initially we 
allow $M_0$ to have an arbitrary curvature, but later we will 
specialise to surfaces of constant curvature. In terms of 
a local complex coordinate $z = x_1 + ix_2$ the metric is
\be
ds_0^2 = \Omega_0(dx_1^2 + dx_2^2) = \Omega_0 \, dz d\z \,,
\ee
where $\Omega_0$ is a position-dependent conformal factor.

The fields we need are a U(1) gauge potential $a$ and a complex Higgs
field $\phi$. Globally, $a$ is a connection on a U(1) line bundle over $M_0$,
and $\phi$ is a section of the bundle. Locally, we represent the
connection as a real 1-form $a = a_1 dx_1 + a_2 dx_2 = 
a_z dz + a_{\z} d\z$. The connection has 2-form field strength 
$f = da$, and we suppose that the first Chern number of the bundle,
\be
N = \frac{1}{2\pi} \int_{M_0} f = \frac{1}{2\pi} \int_{M_0} f_{12} \,
d^2x \,,
\label{Chern}
\ee 
is a positive integer. In component notation $f_{12} = \pr_1 a_2 -
\pr_2 a_1$, and the physical magnetic field strength on $M_0$ is 
$B = \frac{1}{\Omega_0} f_{12}$. More invariantly it is 
$*f$, the Hodge dual of $f$.

The Bogomolny equations are
\bea
D_1\phi + iD_2\phi &=& 0 \,, \label{Bogo01} \\
\frac{1}{\Omega_0}f_{12} &=& 1 - |\phi|^2 \,,
\label{Bogo02}
\eea
where $D_j = \pr_j - ia_j$ is the gauge covariant derivative. They are
also usefully written in terms of the complex coordinate $z$ as
\bea
D_\z \phi &=& 0 \,, \label{Bogo1} \\
-\frac{2i}{\Omega_0} f_{z\z} &=& 1 - |\phi|^2 \,, 
\label{Bogo2}
\eea

The pair of Bogomolny equations (\ref{Bogo1}) and (\ref{Bogo2}) can be
simplified to a single scalar equation as follows. Equation
(\ref{Bogo1}), expanded out, is
\be
\pr_\z \phi - i a_\z \phi = 0 \,,
\ee
and has the solution $a_\z = -i\pr_\z \log\phi$. Since the gauge group is
U(1), $a_z$ is the complex conjugate of $a_\z$, so $a_z = i\pr_z \log\bphi$,
and therefore
\be
f_{z\z} = \pr_z a_\z - \pr_\z a_z = -i\pr_z\pr_\z (\log \phi + \log \bphi) 
= -i\pr_z\pr_\z \log |\phi|^2 \,.
\ee
The second Bogomolny equation (\ref{Bogo2}) therefore reduces to
\be
-\frac{2}{\Omega_0} \pr_z\pr_\z \log |\phi|^2 = 1 - |\phi|^2 \,.
\label{Taubes0}
\ee
It is convenient to change notation, by setting $|\phi|^2 =
e^{2u}$, and to note that the naive Laplacian is $\nabla^2 =
4\pr_z\pr_\z$. Equation (\ref{Taubes0}) then takes the final form
\be
-\frac{1}{\Omega_0} \nabla^2 u = 1 - e^{2u} \,,
\label{Taubes}
\ee
known as the Taubes equation \cite{JT,MS}. The left-hand side is the
negative of the covariant (Beltrami) Laplacian of $u$, and is still
the magnetic field strength.

The first Bogomolny equation (\ref{Bogo1}) implies that $\phi$ is 
gauge-covariantly holomorphic. $\phi$ can therefore have zeros, but 
only of positive multiplicity. These are interpreted as the centres 
of vortices with positive integer winding. It can be shown that the sum of 
the windings around all the vortex centres is
the Chern number $N$. As $\phi$ is zero at each vortex centre, $u$ has
a logarithmic singularity there, and approaches $-\infty$. The Taubes equation
(\ref{Taubes}) is therefore incomplete, and should be supplemented by
delta functions \cite{JT}. We will not include these. Instead, we regard the
Taubes equation as only valid away from the vortex centres, and require
that if there is a vortex centre at $Z$ with winding $n$, then $u$ has
the asymptotic behaviour 
\be
u \sim n \log|z - Z| + {\rm constant}
\label{nearvortex}
\ee
as $z$ approaches $Z$.

Vortex solutions of the Taubes equation should have no further singularities,
so if $M_0$ is compact then $u$ will have a global maximum value, and 
it is interesting to consider the maximum principle in this context. 
The Laplacian of $u$ is non-positive at the location of the global 
maximum of $u$, so the left-hand side of equation (\ref{Taubes}) and 
hence the right-hand side
is non-negative. Therefore $u \le 0$ at the maximum, and hence 
$u \le 0$ everywhere. In known examples, the maximum value of $u$ is
negative. $u$ can have more than one local maximum (typically, between
the vortex centres), and at all of these, $u$ is negative.

If $M_0$ is non-compact, for example the flat plane or the hyperbolic
plane, then we impose the condition $|\phi| = 1$ or equivalently $u =
0$ on the boundary. Again, $u$ is assumed to have no singularities
apart from those at the vortex centres. In this situation, we
also have $u \le 0$ everywhere. If not, then $u$ would have a maximum 
positive value at some point of $M_0$ interior to the boundary. But 
this would again contradict the Taubes equation. Therefore $u$ has 
supremum $0$, attained on the boundary at infinity.

Note that if $u$ is everywhere negative, then the right-hand side of
(\ref{Taubes}) is everywhere positive, so the magnetic field $B$ is
everywhere positive. The magnetic field has its
maximum value $1$ at the vortex centres, where $|\phi|^2 = 0$. 
Physically, the Taubes equation
describes the Meissner effect in a superconductor, where in the
absence of vortices there is no magnetic flux penetration, and
$|\phi| = 1$ everywhere. The vortices introduce magnetic flux defects 
into the superconductor, accompanying the zeros of $\phi$.

\section{More Vortex Equations}

Further vortex equations on $M_0$ arise by changing the coefficients in
equations (\ref{Bogo02}) and (\ref{Bogo2}). The general vortex 
equations we consider are
\bea
D_1\phi + iD_2\phi &=& 0 \label{Bogo01'} \,, \\
\frac{1}{\Omega_0}f_{12} &=& -C_0 + C |\phi|^2 \,,
\label{BogoC}
\eea
or equivalently
\bea
D_\z \phi &=& 0 \,, \label{Bogo1'} \\
-\frac{2i}{\Omega_0} f_{z\z} &=& -C_0 + C |\phi|^2 \,, 
\label{Bogo2'}
\eea
with $C_0$ and $C$ taking any real, constant values. 

As before, we can use the first equation to eliminate the gauge
potential, and by setting $|\phi|^2 = e^{2u}$, the second equation
becomes
\be
-\frac{1}{\Omega_0} \nabla^2 u = -C_0 + C e^{2u} \,.
\label{Scalareq}
\ee
Now, we can simultaneously rescale $C_0$ and $C$ by a positive real 
factor, and absorb this into the metric. We can also rescale $C$
alone by a positive real factor, and absorb this into a constant shift
of $u$ (a rescaling of $|\phi|$). Therefore, without loss of 
generality, we may fix $C_0$ and $C$ to each take one of the three 
standard values $-1$, $0$ or $1$. There are therefore
nine distinct equations of type (\ref{Scalareq}).

Vortex solutions have the property that $\phi$ has zeros, but no  
singularities, and equation (\ref{Bogo1'}) implies that the Chern number
$N$ is positive, because the vortex windings are positive. Of the 
nine equations, only five can have such vortex solutions. The left-hand
side of equation (\ref{Scalareq}) is the magnetic field, and for its
integral to be positive, the right-hand side must admit positive
values. This excludes the four cases $C_0 = 0$ or $C_0 = 1$, combined 
with $C = -1$ or $C = 0$.

The remaining cases are the Taubes equation (\ref{Taubes}), with $C_0 =
C = -1$, the Jackiw--Pi vortex equation \cite{JP1,JP2} with $C_0 = 0$, $C = 1$,
\be
-\frac{1}{\Omega_0} \nabla^2 u = e^{2u} \,,
\label{Jackiw-Pi}
\ee
and the Popov vortex equation \cite{Pop} with $C_0 = C = 1$,    
\be
-\frac{1}{\Omega_0} \nabla^2 u = -1 + e^{2u} \,.
\label{Popov}
\ee
A further case is the equation with $C_0 = -1$, $C = 0$,
\be
-\frac{1}{\Omega_0} \nabla^2 u = 1 \,,
\label{Bradlow}
\ee
that we shall call the Bradlow vortex equation. Notice that the magnetic 
field has constant strength $1$ here. More usually, one refers to the Bradlow 
limit of the Taubes equation for vortices \cite{Bra}. This is where $N$
attains its maximum allowed value on a compact surface, and the Higgs 
field vanishes everywhere. The second Bogomolny equation then says 
that the magnetic field
is $1$, as in equation (\ref{Bradlow}). Our Bradlow vortex
equation is different in that it allows a non-vanishing Higgs field $\phi$
satisfying $D_\z \phi = 0$ in the background of the constant magnetic 
field. Its solutions are therefore similar to what were considered 
previously as vortex solutions close to the Bradlow limit \cite{BM}
or as dissolving vortices \cite{MRom}, where the magnetic field 
was almost constant and the Higgs field small. The final case is the 
vortex equation with $C_0 = -1$, $C = 1$,
\be
-\frac{1}{\Omega_0} \nabla^2 u = 1 + e^{2u} \,.
\label{unnamed}
\ee
This equation, in its flat space version, appeared in Ambj{\o}rn
and Olesen's study of the instability of strong magnetic fields in 
electroweak gauge theory \cite{AO1,AO2}. Notice that the magnetic
field has strength $1$ at the vortex centres, as for Taubes vortices, 
but the strength is enhanced away from these centres. This is an 
anti-Meissner effect.

Applying the maximum principle to the Popov equation (\ref{Popov}), 
we see that $u \ge 0$ at its maximum, and as $u$ approaches $-\infty$ at
the vortex centres, $u$ takes all negative values. For the remaining
equations other than the Taubes equation, the right-hand side is 
positive for all $u$, so there is no further constraint on the value 
of $u$ at its maximum.

\section{Energy and Stability}

The five vortex equations can all be derived using a Bogomolny
rearrangement of a suitable energy functional \cite{Bog}. The energy is not
positive definite in all cases. The equations guarantee
that the energy is stationary, though not always minimal. The vortices
are therefore not necessarily stable.

The energy expression is of the type familiar in the Abelian Higgs
model, but with non-standard coefficients. For general values of $C_0$
and $C$, consider the (potential) energy
\be
E = \int_{M_0} \left\{ \frac{1}{\Omega_0^2} f_{12}^2 
- \frac{2C}{\Omega_0} \left(\overline{D_1 \phi} D_1\phi 
+ \overline{D_2 \phi} D_2\phi \right)  
+ \bigl(-C_0 + C|\phi|^2 \bigr)^2 \right\} \Omega_0 \, d^2x \,.
\ee
This is positive definite if $C \le 0$, but not otherwise. 
We rewrite the energy as
\bea
&& E = \int_{M_0} \bigg\{ 
\left( \frac{1}{\Omega_0}f_{12} + C_0 - C|\phi|^2 \right)^2 \nonumber \\
&& \qquad\qquad\quad
- \frac{2C}{\Omega_0}\bigl(\overline{D_1 \phi} - i\overline{D_2 \phi}\bigr)
\bigl(D_1\phi + iD_2\phi\bigr) \bigg\} \Omega_0 \, d^2x \nonumber \\
&& \quad
+ \int_{M_0} \bigl(-2C_0 f_{12} + 2C f_{12}|\phi|^2 
+ 2Ci\bigl(\overline{D_1 \phi}D_2\phi 
- \overline{D_2 \phi}D_1\phi\bigr)\bigr) \, d^2x \,,
\eea
where the terms in the second integral (which has no $\Omega_0$
factors) compensate for completing the squares. Next
we use the identity
\be
\overline{D_1\phi} D_2\phi - \overline{D_2\phi} D_1\phi = \pr_1(\bphi D_2\phi) 
- \pr_2(\bphi D_1\phi) + if_{12} |\phi|^2 \,,
\ee
which combines the covariant Leibniz rule with the commutator 
$\nobreak{[D_1 , D_2] = -if_{12}}$, to obtain
\bea
E &=& \int_{M_0} \bigg\{ 
\left( \frac{1}{\Omega_0}f_{12} + C_0 - C|\phi|^2 \right)^2 \nonumber \\
&& \qquad\qquad
- \frac{2C}{\Omega_0}\bigl(\overline{D_1 \phi} - i\overline{D_2 \phi}\bigr)
\bigl(D_1\phi + iD_2\phi\bigr) \bigg\} \Omega_0 \, d^2x \nonumber \\
&& 
+ \int_{M_0} \bigl(-2C_0 f_{12}  + 2Ci(\pr_1(\bphi D_2\phi) 
- \pr_2(\bphi D_1\phi))\bigr) \, d^2x \,.
\eea
The final two terms are total derivatives and integrate to zero. More
invariantly, their integral is that of the globally-defined,
exact 2-form $2Ci \, d(\bphi D\phi)$.  
$f_{12}$ integrates to $2\pi$ times the Chern number $N$. Therefore
\bea
E &=& \int_{M_0} \bigg\{ 
\left( \frac{1}{\Omega_0}f_{12} + C_0 - C|\phi|^2 \right)^2 \nonumber \\
&& \qquad\qquad
- \frac{2C}{\Omega_0}\bigl(\overline{D_1 \phi} - i\overline{D_2 \phi}\bigr)
\bigl(D_1\phi + iD_2\phi\bigr) \bigg\} \Omega_0 \, d^2x \nonumber \\
&& - 4\pi C_0 N \,.
\eea
The energy $E$ is stationary and has the value $-4\pi C_0 N$, provided the
Bogomolny equations (\ref{Bogo01'}) and (\ref{BogoC}) are satisfied. 
$E$ is minimised if $C \le 0$. The vortices are then stable, but we are also
interested in cases where $C$ is positive. Taubes and Bradlow vortices
are stable (although the Higgs field does not contribute to
the energy in the Bradlow case, and the equation $D_1\phi + iD_2\phi= 0$ has
to be imposed separately). Popov vortices are unstable, as are
the vortices satisfying equation (\ref{unnamed}). The Jackiw--Pi
vortices are also unstable by this criterion, but this is not of much
significance, as these vortices arise most naturally in
Chern--Simons field theory, where the dynamics is different \cite{JP2,Dun}.

Euler--Lagrange equations can be derived from the energy $E$; these
are the second-order, static field equations for vortices. They are 
satisfied for all types of vortices satisfying the appropriate
first-order Bogomolny equations, because $E$ is stationary. This can 
be checked by differentiation.

\section{Vortices as Conical Geometry}

The original metric $ds_0^2$ on $M_0$ is smooth, but it was
suggested some time ago \cite{CM} that it is useful to consider
for a vortex solution the metric
\be
ds^2 = e^{2u} \, ds_0^2 \,,
\label{Bapmetric}
\ee
the original metric conformally rescaled by the squared Higgs field 
$\nobreak{|\phi|^2 = e^{2u}}$. This modified metric has been studied in depth for Abelian
Higgs vortices by Baptista \cite{Bap}, so we refer to it as the Baptista
metric. $ds^2$ defines an intrinsic
geometry of a vortex solution, and it is sometimes easier to
describe and visualise this intrinsic geometry, rather than separate
$ds^2$ into $ds_0^2$ and $e^{2u}$. The Baptista metric is useful for
all five of our vortex equations, although it has different properties in
the various cases. Notice that it tends to reduce 
lengths and areas near vortex centres, because $e^{2u}$ is close to zero.
 
The Baptista metric is not a regular Riemannian metric, because it
vanishes at the vortex centres. Taubes showed that the asymptotic form of
$u$ near a vortex centre is as in equation (\ref{nearvortex}). For a 
vortex with unit winding ($n=1$), centred at the origin $Z=0$ for convenience,
$e^{2u} \sim \mu|z|^2$ with $\mu$ a positive constant. The background
metric is locally $\Omega_0(0) dz d\z$ with $\Omega_0(0)$ positive, so
the Baptista metric is locally $\mu\Omega_0(0)|z|^2 dz d\z$. In polar 
coordinates, this is a multiple of $r^2(dr^2 + r^2d\theta^2)$. 
Using the change of variables $\rho = \half r^2$ and $\chi = 2\theta$, 
the metric becomes $d\rho^2 + \rho^2 d\chi^2$, a flat metric whose polar
angle $\chi$ runs from $0$ to $4\pi$. The metric is therefore conical, 
with cone angle $4\pi$. The conical excess is $2\pi$. For a vortex centre of
multiplicity $n$, the cone angle would be $2(n+1)\pi$, with excess
$2n\pi$. The Baptista metric is not truly a flat cone, because there
are higher-order metric corrections, and generally there is a non-zero
curvature as the conical singularity is approached.

Baptista derived a simple relation between the curvature of the background
metric and the curvature of the new (Baptista) metric. We present this for the
general vortex equation $-\frac{1}{\Omega_0} \nabla^2 u = -C_0 + C e^{2u}$.
We start with the formula for the Gaussian curvature of the background,
\be
K_0 = -\frac{1}{2\Omega_0} \nabla^2 \log \Omega_0 \,.
\label{GaussianCurv}
\ee
The Baptista metric, with conformal factor $\Omega = e^{2u} \Omega_0$, has
Gaussian curvature
\be
K = -\frac{1}{2e^{2u}\Omega_0} \nabla^2(2u + \log \Omega_0) \,,
\ee
so the curvatures are related by
\be
-\frac{1}{\Omega_0}\nabla^2 u = -K_0 + Ke^{2u} \,.
\ee
This is a well known, purely geometrical identity \cite{Ber,KW},
discussed in the context of metrics with conical singularities by
Troyanov \cite{Tro}. In addition, $u$ satisfies the vortex equation, so
\be
-C_0 + C e^{2u} = -\frac{1}{\Omega_0}\nabla^2 u = -K_0 + Ke^{2u} \,.
\ee

Baptista's version of this equation, obtained by multiplying by $\Omega_0$, is
\be
(K_0 - C_0)\Omega_0 = (K-C)\Omega \,.
\label{Baprel}
\ee
Intrinsically, this relates linear combinations of the curvature
2-form and K\"ahler 2-form of the background metric and Baptista metric.  
From it, Baptista derived a superposition principle for Taubes vortices
\cite{Bap}. Equation (\ref{Baprel}) is not algebraic, despite its
appearance, because the curvature formulae involve the Laplacian.

A basic property of the Baptista metric is the relation between its area 
\be
A = \int_{M_0} \Omega \, d^2x 
\ee
and the background area of $M_0$, 
\be
A_0 = \int_{M_0} \Omega_0 \, d^2x \,.
\ee
If $M_0$ is smooth and compact, and of genus $\mg_0$, then by the 
Gauss--Bonnet theorem,
\be
\frac{1}{2\pi} \int_{M_0} K_0 \, \Omega_0 \, d^2x = 2 - 2\mg_0 \,.
\ee
For an $N$-vortex solution, the curvature of the Baptista metric 
on $M_0$ satisfies  
\be
\frac{1}{2\pi} \int_{M_0} K \, \Omega \, d^2x \, - N = 2 - 2\mg_0 \,,
\ee
because each conical singularity with conical excess $2\pi$
contributes $-1$ to the Gauss--Bonnet integral, and the topology of
$M_0$ is unchanged. Integrating the equation (\ref{Baprel}),
and using these formulae, we find
\be
CA = C_0 A_0 + 2\pi N \,.
\ee

This has the following consequences for the five vortex equations in 
standard form. For the Taubes equation with $C_0 = C = -1$, 
$A = A_0 - 2\pi N$. The Baptista area $A$ is smaller than the 
original area $A_0$, implying Bradlow's upper bound on the vortex number 
$2\pi N \le A_0$, because $A$ has to be non-negative. It is not possible 
to have more vortices than this satisfying the Bogomolny equations. 
If $A_0 = 2\pi N_0$ for some integer $N_0$, and $N=N_0$,
then we are at the Bradlow limit of the Taubes equation. Here $A=0$, 
so the Higgs field and the Baptista metric are both zero. Genuine 
Taubes vortices require $N < N_0$. 

Our Bradlow vortex equation (\ref{Bradlow}), with $C_0 = -1$ and
$C = 0$, allows for vortices in this limit. Solutions only exist if 
$A_0 = 2\pi N$ (as the magnetic field strength is $1$, so its 
integral is $A_0$). The Higgs field satisfies $D_\z \phi = 0$ in the
background of the constant magnetic field, which is the equation for
Lowest Landau Level states. $\phi$ can 
be non-zero, and its magnitude can be rescaled by an arbitrary
constant, so the area $A$ is arbitrary.  
 
For the Popov vortices with $C_0 = C = 1$, $A$ is larger than $A_0$
and there is no constraint on $N$ (although we shall see later that
$N$ must be even). This implies that the average of $e^{2u}$ over
$M_0$ is greater than $1$, so $u$ must be strictly positive at its 
maximum, and take all values between this maximum and $-\infty$, a 
stronger result than what we obtained using the maximum principle. 
For the Jackiw--Pi vortices with $C_0 = 0$ and 
$C = 1$, the Baptista metric has area $A = 2\pi N$. 

For the vortices satisfying equation (\ref{unnamed}), with $C_0 = -1$ and 
$C = 1$, the vortex number has to satisfy $2\pi N > A_0$ for $A$ to be 
positive. This is a novel lower bound on the vortex number. If 
$2\pi N = A_0$ we again have a Bradlow limit, with degenerating vortices.

\section{Integrable Vortices}

When the curvature $K_0$ of the background surface $M_0$ is constant, with a
special value adapted to the vortex equation, we call the vortex equation
integrable. The special curvature values are those that make each side
of the Baptista equation (\ref{Baprel}) vanish. The vortex equation is
therefore integrable if $K_0 = C_0$. So for the Taubes, Bradlow and final type 
of vortices, with standard coefficients, $M_0$ needs to be hyperbolic, with 
curvature $K_0 = -1$. For the Jackiw--Pi vortices the background needs to 
be flat, with $K_0 = 0$, and for the Popov vortices the background
needs to be spherical, with $K_0 = 1$. The Baptista metric then has
constant curvature $K = C$, except at the conical singularities, so 
in these five cases it is respectively hyperbolic, flat, spherical, 
spherical and spherical.

Finding the geometry of these integrable vortices is closely related
to the problem of starting with a smooth, constant curvature Riemann surface 
$M_0$ and constructing on it another constant curvature metric (possibly
with different curvature) that additionally has a number of
conical singularities, each with conical excess $2\pi$. This
requires solving Liouville's equation. Solutions of Liouville's
equation can locally be expressed in terms of a holomorphic function 
$f$, and the conical singularities correspond to ramification points 
of $f$, where the derivative of $f$ vanishes.   

For some of the equations we are discussing, this construction of
vortex solutions is well known. We briefly review these cases. Then we
discuss cases that have not been considered before, and find a
few novel solutions. 

The reduction to Liouville's equation is simple. Let us write the
Baptista conformal factor as $\Omega = e^{2v}$. Then 
the curvature formula $K = - \frac{1}{2\Omega}
\nabla^2 \log\Omega$ becomes $\nabla^2 v = -Ke^{2v}$, and therefore
\be
\nabla^2 v = -C e^{2v}
\label{Liouville}
\ee
when $K = C$. This is Liouville's equation when $C$ is non-zero,
but the Bradlow case $C=0$ can be included too. 

The general solution of equation (\ref{Liouville}) 
in a simply connected region of the $z$-plane is
\be
\Omega = e^{2v} 
= \frac{4}{(1 + C|f(z)|^2)^2} \left| \frac{df}{dz} \right|^2 \,,
\ee
where $f$ is a holomorphic function. This formula is also valid if
$C=0$, as $v$ is then the sum of the holomorphic function 
$\half\log\left(2 \frac{df}{dz}\right)$ 
and its complex conjugate, and therefore satisfies Laplace's
equation. Locally we also have an explicit expression for the
background conformal factor $\Omega_0$. In suitably chosen local coordinates,
\be
\Omega_0 = \frac{4}{(1 + C_0|z|^2)^2} \,.
\ee
The solution of all the integrable vortex equations is therefore locally
\be
|\phi|^2 = e^{2u} = \frac{\Omega}{\Omega_0} 
= \frac{(1 + C_0 |z|^2)^2}{(1 + C|f(z)|^2)^2} 
\left| \frac{df}{dz} \right|^2 \,,
\label{SquHiggsfld}
\ee
with the values of $C_0$ and $C$ as in the vortex equation. 

One may fix the gauge by choosing the Higgs field itself to be
\be
\phi = \frac{1 + C_0 |z|^2}{1 + C |f(z)|^2} \frac{df}{dz} \,.
\label{Higgsfld}
\ee
The vortex centres are the ramification points of $f$, where its 
derivative vanishes. Here, $\phi$ is zero. If $f$ near $Z_0$ has the
expansion 
\be
f(z) \sim f_0 + \nu(z-Z_0)^{n+1}
\ee
then the ramification number is $n$, and there are $n$ coincident 
vortices at $Z_0$.

The original application of these formulae was Witten's construction of
Taubes vortices on the hyperbolic plane \cite{Wit}. Here, $C_0=C=-1$ 
and $f$ is a holomophic map from the hyperbolic plane to itself, 
mapping boundary to boundary. In the Poincar\'e disc model, $f$ needs
to be a Blaschke rational function
\be
f(z) = \prod_{m=1}^{N+1} \frac{z - a_m}{1 - \overline{a_m}z}
\ee
with $|a_m| < 1$. Inside the disc, $\frac{df}{dz}$ has $N$ zeros, so
there are $N$ vortices, and the Higgs field satisfies the boundary condition
$|\phi| = 1$. The simplest example is where $f(z) = z^{N+1}$, and the 
expression (\ref{Higgsfld}) for the Higgs field is
\be
\phi = \frac{1 - |z|^2}{1 - |z|^{2(N+1)}} (N+1)z^N 
= \frac{(N+1)z^N}{1 + |z|^2 + \cdots + |z|^{2N}} \,.
\ee
Here, there are $N$ coincident vortices at the origin. 

Similar formulae have been used to construct solutions of the
Popov vortex equation \cite{Man}, with $C_0=C=1$. These are vortices 
on a unit sphere, and are constructed using a meromorphic function 
$f$, a map from the Riemann sphere to itself. To obtain a finite 
vortex number, the map must again be rational, of the form
\be
f(z) = \frac{p(z)}{q(z)} \,,
\ee
where $p$ and $q$ are any polynomials with no common root. If $p$ and
$q$ (and hence $f$) have degree $n$, then $\frac{df}{dz}$ has $2n-2$ zeros. The
vortex number for Popov vortices is therefore an even number, 
$N = 2n-2$. It has been shown by Chen et al. \cite{Che} that this 
construction of integrable Popov vortices, using global rational 
functions, gives all solutions.

The Riemann--Hurwitz formula states that if $f$ is a globally-defined 
holomorphic map of degree $n$ from a compact surface of genus $\mg_0$ to a 
compact surface of genus $\mg$, then the ramification number (vortex 
number) $N$ is given by
\be
2 - 2\mg_0 + N = n(2 - 2\mg) \,.
\label{RiemHur}
\ee
If the map is from a sphere to a sphere then $\mg_0 = \mg = 0$, so
$N = 2n-2$, confirming the vortex number given above.

Also known are solutions of the Jackiw--Pi vortex equation, both 
on the flat plane, and on a torus. Let us focus on solutions
on a torus \cite{HZ}, which have the form (\ref{SquHiggsfld}) with 
$C_0=0$ and $C=1$,
\be
|\phi|^2 = \frac{1}{(1 + |f(z)|^2)^2} \left| \frac{df}{dz} \right|^2 \,.
\ee
The simplest solutions are
where $f$ is a globally-defined holomorphic map from the torus to 
a sphere, i.e. a meromorphic function on the torus. $f$ is then 
an elliptic function, with the double periodicity of the torus.
For an elliptic function, $\mg_0 = 1$ and $\mg = 0$, and the degree $n$ 
is the number of poles (counted with multiplicity). The vortex number 
is then $N = 2n$ according to the Riemann--Hurwitz formula. The 
simplest elliptic functions have degree $2$ and give solutions with 
four vortices. 

$f$ may not be globally defined, and there is a Jackiw--Pi vortex
solution where $f$ is elliptic but the periods of $f$ are twice the 
periods of the torus itself. This solution is due to Olesen \cite{Ole}, 
and has vortex number $N=1$. It has been presented in slightly simpler
form, and given a braneworld interpretation, in \cite{AC}. 
When the torus is glued together, $f$
transforms, but in such a way that $|\phi|^2$ is smooth. It would be useful
to more systematically investigate vortices constructed from
local holomorphic maps $f$, suitably glued together.  

A few Taubes vortex solutions are known on hyperbolic surfaces other than the
hyperbolic plane \cite{MR}, and in particular on the Bolza surface
\cite{MM}, the most symmetric genus 2 surface. The Bolza surface is a 
double covering of the Riemann sphere, branched over the six
vertices of a regular octahedron. Algebraically, it is defined as the
complex curve
\be
y^2 = (x^4 - 1)x  \quad (x,y \in \C) \,,
\label{Bolzacurve}
\ee
with branch points at $0,1,i,-1,-i$ supplemented by a branch 
point at infinity. The Bolza surface has a hyperbolic metric 
in which the octants of the Riemann sphere 
are covered by 16 equilateral, hyperbolic triangles with vertex 
angles $\frac{\pi}{4}$. At each branch point, eight of these 
triangles meet smoothly. The Bolza surface can also be modelled as a 
regular hyperbolic octagon with vertex angles $\frac{\pi}{4}$ and 
opposite sides identified, and all eight vertices identified to one 
point. This octagon is one cell of the $\{8,8\}$ tesselation of the 
Poincar\'e disc by octagons.

A compact hyperbolic surface of curvature $-1$ has area 
$A_0 = 2\pi(2\mg_0 - 2)$, by the Gauss--Bonnet theorem. Therefore, for the 
(integrable) Taubes vortex equation, $N < 2\mg_0 - 2$, and there can 
be no more than one vortex on the $\mg_0 = 2$ Bolza surface. An 
explicit solution has been found for a vortex located at a branch
point (by symmetry, all branch points are equivalent) \cite{MM}. In 
the octagon model, two of the branch points correspond to the centre of the 
octagon and the vertex of the octagon, and the solution with a vortex
at either point is available. The formulae depend on a compound
Schwarz triangle map $f$, which is locally a map from the hyperbolic 
plane to itself, mapping a triangle to a triangle. One triangle angle 
is doubled by the map, and this produces the ramification of $f$ 
required at the vortex centre.

The solutions considered so far were all previously known. Let us
briefly mention a class of solutions of the vortex equation 
(\ref{unnamed}), with $C_0 = -1$, $C = 1$. This equation is integrable 
on a hyperbolic surface $M_0$ with $K_0 = -1$, and 
requires $f$ to be locally a map from $M_0$ to a sphere. 
If $f$ is globally defined, then it is a meromorphic
function on $M_0$. Meromorphic functions on compact surfaces 
are plentiful, although not generally easy to write down explicitly.
For the Bolza surface defined by equation (\ref{Bolzacurve}), the 
simplest meromorphic function is $x$. The map 
$f$ is then the canonical covering map from the Bolza surface to the
Riemann sphere, of degree 2. $f$ has six ramification points, according
to the Riemann--Hurwitz formula (\ref{RiemHur}), and they are the
lifts of the six branch points on the sphere. For example, in the 
neighbourhood of $x=0$, $y$ is a good local coordinate and 
$x = y^2 + \cdots$. So $f(y) = y^2 + \cdots$ and $\frac{df}{dy} = 0$ 
at $y=0$. The function $x$ therefore gives a 6-vortex solution of
equation (\ref{unnamed}) on the Bolza surface, a vortex number 
consistent with the inequality $N > 2\mg_0 - 2$. The Baptista metric 
is simply the sphere metric $\frac{4}{(1 + |x|^2)^2} \, dxd{\bar x}$ 
lifted to the double cover. One would need to express $x$ in terms 
of $y$ to make this lift explicit, which requires solving a quintic.

Integrable Bradlow vortices on a hyperbolic surface $M_0$ locally 
involve maps from $M_0$ to a flat surface. The Baptista metric is 
then flat, apart from its conical singularities. Such metrics arise 
from Abelian differentials of the first kind (holomorphic 1-forms), as
follows. Given any such differential form $\omega$ on $M_0$, there is the 
beautifully simple metric 
\be
ds^2 = |\omega|^2 \,.
\ee
Because $\omega$ is closed it can be expressed locally as 
$\omega = \frac{df}{dz} \, dz$, so the metric is 
\be
ds^2 = \left| \frac{df}{dz} \right|^2 dzd\z \,, 
\ee
which is flat, being the pullback of the flat metric $dw d\w$ using the
map $w = f(z)$. The metric also has conical singularities where
$\omega = 0$, or equivalently at the ramification points of
$f$. 

Globally, an Abelian differential of the first kind is a section
of the canonical bundle, and has $2\mg_0 - 2$ zeros, where $\mg_0$ is the genus
of $M_0$. The vortex number is therefore $N = 2\mg_0 - 2$, as expected for
Bradlow vortices.  We will describe more explicitly a Bradlow
vortex solution on the $\mg_0 = 2$ Bolza surface in the next section.

\section{Geometric Interpretation of Vortices}

Associated to a vortex solution, there is the conformal modification 
of the background metric, which we are calling the Baptista metric. Its 
curvature satisfies equation (\ref{Baprel}), but note that this is 
not easy to solve on a general surface,
and is not equivalent to the problem of constructing a metric with 
given curvature. The exceptions are the cases where the vortex equation
is integrable. Here, finding a vortex solution is equivalent to finding
a metric of constant curvature, with conical singularities
of cone angle $4\pi$ at specified vortex locations. 
For some purposes, one may regard the Baptista metric 
as an intrinsic geometry of a vortex, and in this section, we shall 
explore this intrinsic geometry further.

Recall that the definition of the Baptista metric (\ref{Bapmetric}) 
implies that the squared Higgs field of a vortex solution on $M_0$ is the 
ratio of two conformally equivalent metrics,
\be
|\phi|^2 = e^{2u} = \frac{ds^2}{ds_0^2} = \frac{\Omega}{\Omega_0} \,.
\ee
For integrable vortices, this is the ratio of a constant 
curvature metric on $M_0$ with conical singularities at the vortex 
locations (the Baptista metric with curvature $C$) to the smooth, 
constant curvature metric on $M_0$ (the background metric with
curvature $C_0$).

We can verify directly that $|\phi|^2$ satisfies the vortex equation
(\ref{Scalareq}). Let us write $\Omega_0 = e^{2t}$ and $\Omega =
e^{2v}$, so that
\be
e^{2u} = e^{2(v-t)}
\ee
and therefore $u = v-t$. The general formula (\ref{GaussianCurv}) for the
curvature implies that $\nabla^2 t = -C_0 e^{2t}$, and similarly 
$\nabla^2 v = -C e^{2v}$. Therefore, 
\be
\nabla^2 u = -C e^{2v} + C_0 e^{2t} \,,
\ee
and dividing by $e^{2t}$ we obtain $e^{-2t} \nabla^2 u = - C e^{2u} + C_0$,
which is equivalent to (\ref{Scalareq}).

This geometric description of a vortex solution matches the formula 
(\ref{SquHiggsfld}) as follows. The background metric on $M_0$ has 
constant curvature $C_0$, so is locally 
\be
ds_0^2 = \frac{4}{(1 + C_0 |z|^2)^2} \, dzd\z \,.
\label{backgrdmet}
\ee
$f$ is a holomorphic map (at least locally) from $M_0$ to a second constant
curvature Riemann surface $M$ with curvature $C$. Let this surface 
have local complex coordinate $w$ and metric
\be
\widetilde{ds^2} = \frac{4}{(1 + C |w|^2)^2} \, dwd\w \,.
\ee
The map has the local expression $w = f(z)$, so 
$dw = \frac{df}{dz} dz$. The metric $\widetilde{ds^2}$, pulled 
back to $M_0$ using the map $f$, is therefore
\be
ds^2 = \frac{4}{(1 + C |f(z)|^2)^2} \left| \frac{df}{dz} \right|^2 \,
dzd\z \,,
\label{Bapmet}
\ee
and this is the Baptista metric. It still has curvature $C$, but
also has conical singularities at the ramification points of $f$, the
locations of the vortices. The ratio of the metrics (\ref{Bapmet}) 
and (\ref{backgrdmet}) then gives the formula (\ref{SquHiggsfld}).

The Baptista metric, being the pullback of a constant curvature 
metric, can sometimes be described explicitly. For example, it 
was shown in \cite{MM} that an $N=1$ Taubes vortex on the Bolza
surface, located at the centre of
the hyperbolic Bolza octagon, is obtained using a map $f$ from the octagon to 
a hyperbolic square with the same vertex angles, $\frac{\pi}{4}$. The 
map double covers the square, with a branch point at the centre of 
the square. The full image winds round the square twice and is 
therefore itself an octagon with a conical singularity at the centre. 
Its hyperbolic metric, including the 
singularity, is the Baptista metric. The pullback by $f$ places 
this metric on the original octagon. Opposite sides are identified 
in the same way for both octagons. The octagon with
the Baptista metric has half the area of the original Bolza octagon.
(Note that $f$ is not a globally-defined map between compact surfaces,
because the hyperbolic square by itself does not have opposite sides
identified. The appropriate modification of equation (\ref{RiemHur})
is discussed in \cite{MM}.) 

It is rather easier to visualise this geometry if we consider the
vortex to be at the vertex of the Bolza octagon, which is equivalent
by symmetry to being at the centre. In this case the
appropriate map $f$ is from the Bolza octagon with vertex angles
$\frac{\pi}{4}$ to a smaller hyperbolic octagon with vertex angles
$\frac{\pi}{2}$. Gluing opposite sides of the smaller octagon together
creates a cone of angle $4\pi$ at the vertex, as required. This glued-together
octagon is the intrinsic geometry of the vortex with its Baptista 
metric, and some of its geometric properties are easy to calculate. 
This is despite the fact that the map $f$ involves Schwarz triangle 
functions (and hence hypergeometric functions), so is not very explicit. 

\begin{figure}
\centering
\includegraphics[width=5in]{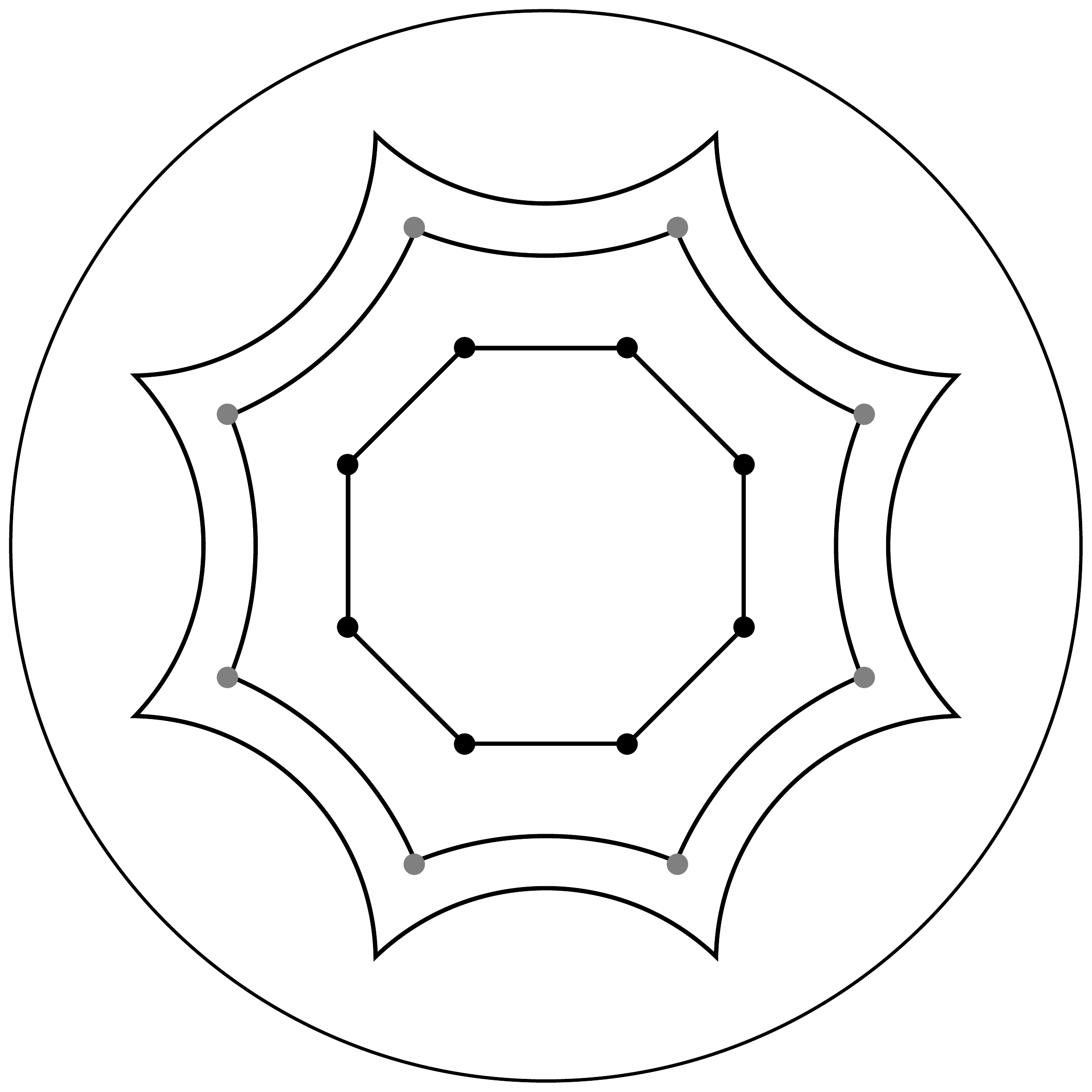}
\vskip 5pt
\caption{Three octagons: The outer is the Bolza octagon with vertex
angle $\frac{\pi}{4}$ superimposed on the Poincar\'e disc; the middle 
is the Baptista octagon with vertex angle $\frac{\pi}{2}$ -- the intrinsic 
(hyperbolic) geometry of an $N=1$ Taubes vortex located at the Bolza octagon 
vertex; the inner is a flat octagon with vertex angle $\frac{3\pi}{4}$ -- 
the intrinsic geometry of an $N=2$ Bradlow vortex located at the 
Bolza octagon vertex. In all cases, opposite edges are identified.}
\label{Vortoctagons}
\end{figure}

These two octagons are shown in Figure \ref{Vortoctagons}. The 
intrinsic, Baptista octagon has half the area of the Bolza 
octagon. Using hyperbolic trigonometry, one can compare the
lengths of their closed geodesics, for example, the geodesics along
the boundary connecting vertex to vertex. Assuming the curvature is
$K_0 = -1$, the boundary geodesic of the Bolza octagon has length $a$,
where $\cosh a = 5 + 4\sqrt{2}$. The analogous geodesic on the octagon
with its Baptista metric and conical singularity has the
shorter length $\tilde a$, where $\cosh {\tilde a} = 1 + \sqrt{2}$.

An $N=2$ Taubes vortex on the Bolza octagon would saturate the Bradlow 
limit, and the Baptista metric would degenerate and have zero area. 
We have evaded this degeneracy by introducing the Bradlow vortex equation 
(\ref{Bradlow}). For this equation on the Bolza surface, the vortex number 
$N$ must be 2. The solution for the squared Higgs field is given by 
equation (\ref{SquHiggsfld}) with $C=0$. The Baptista metric is 
therefore the pullback of a flat metric to the Bolza surface
\be
ds^2 = \left| \frac{df}{dz} \right|^2 \, dzd\z \,.
\ee
For a vortex of multiplicity 2 at the Bolza octagon vertex, the
intrinsic geometry is now a flat, regular octagon of arbitrary scale 
size, with its straight, opposite sides identified. This octagon is 
also shown in Figure \ref{Vortoctagons}. The
vertex angle is $\frac{3\pi}{4}$, so after gluing, there is a single
conical singularity of cone angle $6\pi$. The conical excess is
$4\pi$, as expected for a vortex of multiplicity 2. The map $f$ is
from the hyperbolic Bolza octagon to the flat octagon, which again
involves nontrivial Schwarz triangle maps. The intrinsic geometry of
the vortex is fairly easy to visualise, even though the squared Higgs field
is not an elementary function.

The intrinsic geometry of Popov vortices is a spherical metric on a
unit-sphere background, incorporating conical singularities. For example, 
the 2-vortex intrinsic geometry is a double covering of the unit sphere
branched over a pair of points, with the unit-sphere metric lifted to
both sheets. The total area is therefore $8\pi$, twice the original
area. On the background sphere, the vortex locations can be at any
pair of distinct points. Coincident vortices are not allowed, because
a rational map of degree 2 cannot have a single ramification point. 
It is also known that in the intrinsic spherical geometry with two conical
singularities, the cone angles must be equal, and the singularities
are at antipodal points \cite{Tro2}. This is intuitively fairly
clear. A conical singularity (with an angular deficit or angular 
excess) at the north pole opens up a wedge bounded by meridians (geodesics), 
and these meet at the south pole at the same angle. Then these
meridians are glued together. The Baptista metric for two Popov 
vortices must therefore have this geometry with cone angles $4\pi$, 
for any rational map of degree 2. 

It is also easy to describe examples of the intrinsic geometry of 
the Jackiw--Pi vortices, and of the vortex solutions of equation 
(\ref{unnamed}). The background surface should be a flat 
torus of genus $\mg_0 = 1$ in the Jackiw--Pi case, and a hyperbolic 
surface of higher genus $\mg_0$ in the second case. Suppose the 
background surface is hyperelliptic, a double cover of a sphere 
with $2\mg_0 + 2$ branch points. (This always holds for surfaces of 
genus $1$ or $2$, but only for selected surfaces of higher genus.) 
A special vortex solution is then obtained by choosing $f$ to be the
covering map. The vortex number is $N = 2\mg_0 + 2$ and the vortices 
are located at the ramification points (the points covering the 
branch points). The Baptista metric is the underlying spherical 
metric pulled back to the double cover, and it has conical
singularities with cone angle $4\pi$ over each branch point. 
$|\phi|^2$ is the ratio of this lifted spherical metric to the smooth, 
background flat or hyperbolic metric, but is not an elementary function. 

We can go beyond double covers. Suppose $f:M_0 \to {\tilde M}$ is any 
globally-defined, branched covering map between compact Riemann 
surfaces. Both $M_0$ and $\tilde M$ have unique constant curvature 
metrics. The pullback by $f$ of the metric on $\tilde M$ to $M_0$ is 
a Baptista metric $ds^2$ of vortices on $M_0$, with vortices located 
at the ramification points of $f$, and its ratio to the smooth metric 
$ds_0^2$ on $M_0$ is the squared Higgs field. The relevant vortex 
equation that is satisfied depends on the two curvatures.

\section{Conclusions}

We have considered the generalised Taubes equation for vortices on
a curved background surface,
$-\frac{1}{\Omega_0} \nabla^2 u = -C_0 + C e^{2u}$. By rescaling, both
$C_0$ and $C$ take standard values $-1$, $0$ or $1$, but only five
combinations of these values allow vortex solutions without
singularities. After reviewing Baptista's \linebreak metric $ds^2 = |\phi|^2
ds_0^2$, where $ds_0^2$ is the background metric and $\phi$ the Higgs 
field, we have seen that the vortex equation is integrable provided 
the background curvature is constant and equals $C_0$. Baptista's 
metric is then of constant curvature $C$, but additionally, for an
$N$-vortex solution, it has $N$
conical singularities with cone angles $4\pi$. Solutions of Liouville's
equation, locally involving a holomorphic function $f$, give 
constant curvature metrics, and the conical singularities arise from
ramification points of $f$. This allows a unified treatment of known 
solutions for Taubes, Jackiw--Pi and Popov vortices, and also for the two
further types of vortex presented here, in all the integrable cases. The
squared Higgs field on a compact Riemann surface $M_0$ is simply the 
ratio of a constant curvature metric with conical singularities 
to the unique, smooth constant curvature metric. 

For some vortex solutions,
including 1- and 2-vortex solutions on the genus-2 Bolza surface, we
have described the intrinsic Baptista geometry, bypassing the need
for finding the Higgs field explicitly. It would be desirable to
extend this intrinsic geometrical picture of vortices to further examples. 

For integrable vortices, the Baptista metric with its constant curvature
and conical singularities is the spatial part of an Einstein metric
in 2+1 dimensions with cosmological constant \cite{DJtH,DJ}. The singularities 
have a $2\pi$ conical excess, and therefore correspond to point-particle 
sources of negative mass. It is a surprise that vortices, which are 
normally regarded as smooth field configurations on a smooth surface, 
have such a point-particle interpretation, and it would be interesting 
if the dynamics of vortices could be related to the dynamics of 
gravitating point particles.   

The integrable cases of the Taubes and Popov vortex equations, where the
background geometry is respectively hyperbolic and spherical, are
known to arise from a dimensional reduction of the self-dual Yang--Mills 
equations in $\R^4$ \cite{Wit,Pop}. Vortex solutions can therefore be
interpreted as instantons with symmetry. A more systematic treatment of the
dimensional reduction, allowing for a wider range of symmetry groups and 
gauge groups, can probably account for all the vortex equations considered 
here. This is under investigation by Contatto and Dunajski \cite{CD}. 

\section*{Acknowledgements}

I am grateful to the organisers of the LMS/EPSRC Durham symposium on
Geometric and Algebraic Aspects of Integrability (August 2016). This 
work was partially completed during the symposium, and I thank 
Robert Bryant in particular for helpful remarks. I also warmly 
thank Rafael Maldonado for comments, and for producing Figure 1.

\end{document}